\documentclass[11pt]{article}

\input epsf.sty

\def\lromn#1{\uppercase\expandafter{\romannumeral#1}}

\usepackage{graphicx,amsmath,amssymb}
\begin{document}

\begin{center}
\begin{Large}
\textbf{
Neutrino mass spectroscopy using Er$^{3+}$ ions placed at inversion center
of host crystals
}
\end{Large}

\vspace{1cm}

H. Hara, N. Sasao, A. Yoshimi, and M. Yoshimura

Research Institute for Interdisciplinary Science,
Okayama University \\
Tsushima-naka 3-1-1 Kita-ku Okayama
700-8530 Japan

\vspace{7cm}

\begin{Large}
{\bf ABSTRACT}
\end{Large}
\end{center}

We propose neutrino mass spectroscopy  using Er$^{3+}$:Cs$_2$NaYF$_6$ or :Y$_2$O$_3$ crystal placed in hollow of a
Bragg fiber as a target system.
Unknown neutrino parameters and properties such as the lightest neutrino mass,
Majorana/Dirac distinction, and CP violating phases can
be explored by measuring  scattered  photons ($\gamma$) along
the excitation (and fiber) axis by varying Raman trigger ($\gamma_0$) directions,
in Er$^{3+}$ de-excitation process from $|e\rangle $ state
to $|g\rangle $ state; $|e\rangle \,, | e\rangle + \gamma_0 \rightarrow | g\rangle + \gamma + \nu_i\bar{\nu}_j$,
$\nu_i\,, i = 1, 2,3$ being a mass-resolved neutrino state.
Rates and required level of QED background rejection are calculated using measured data of the target system.

\vspace{5cm}
Keywords
\hspace{0.5cm} 
Neutrino mass,
Majorana fermion,
lanthanoid ions at inversion center,
Er$^{3+}$,
Bragg fiber

\newpage

\section
{\bf Introduction}

In two preceding papers \cite{ranp}, \cite{ranp 2} some of us explored the possibility of
using lanthanoid ions doped in crystals in order to open a pathway towards
resolving remaining important issues of neutrino properties:
(1) the smallest neutrino mass in the three flavor neutrino scheme,
(2) Majorana/Dirac distinction along with their CP violating phases.
The purpose of the present work is to investigate schemes using  promising
target ion Er$^{3+}$ for the same objectives.
Er$^{3+}$ ions placed at inversion center of host crystals 
such as Cs$_2$NaYF$_6$ and Y$_2$O$_3$ are attractive candidates of neutrino mass spectroscopy 
\cite{renp overview},
since 100 \% Er  replacing  Y site is possible, hence a high density target may
be realized.
Moreover, this ion is considerably different from Sm$^{2+}$ previously studied \cite{ranp 2} 
in the energy level structure,
and above all it is a Kramers ion for which a magnetic control may work in interesting manners.

Experimental progress towards neutrino mass spectroscopy \cite{renp overview} relies on
the principle of macro-coherence which makes otherwise tiny rates large enough of
experimental interest.  Its verification in weak QED processes has been demonstrated in \cite{psr exp},
realizing $\sim 10^{18}$ effective enhancement over spontaneous emission rate.
The next important step is to fabricate solid targets, and for this purpose lanthanoid ions are
ideal due to its narrow optical width of state levels.
Reduction of macro-coherently amplified QED backgrounds is the main issue of discussion, since rates
can be made large enough for detection.
The largest QED background is rejected by a kinetic constraint and the next largest by
a symmetry of host crystals, presence of inversion center \cite{ranp 2}.
We employ  Bragg fiber 
\cite{renp in bragg fiber}, \cite{renp in bragg fiber 2} to reject still remaining backgrounds.

We use the natural unit of $\hbar = c = 1$ throughout the present work unless otherwise stated.
Useful numbers to remember are 1 eV$ = 1.5 \times 10^{15}$sec$^{-1}$ and its inverse $= 1,240$ nm of laser wavelength ,
Avogadro number cm$^{-3} = 7.6 \times 10^{-15}  $eV$^{3}$, $G_F^2\,$eV$^5 = 2.1 \times 10^{-31} $sec$^{-1}$.
Atomic physics often uses a unit of energy, cm$^{-1}$, and it is related to eV by $10^{4}$cm$^{-1} = 1.24$eV.

\vspace{0.5cm} 
\section
{\bf Three RANP schemes and rejection of major QED background}

\begin{figure*}[htbp]
 \begin{center}
 \epsfxsize=0.6\textwidth
 \centerline{\includegraphics[width=17cm,keepaspectratio]{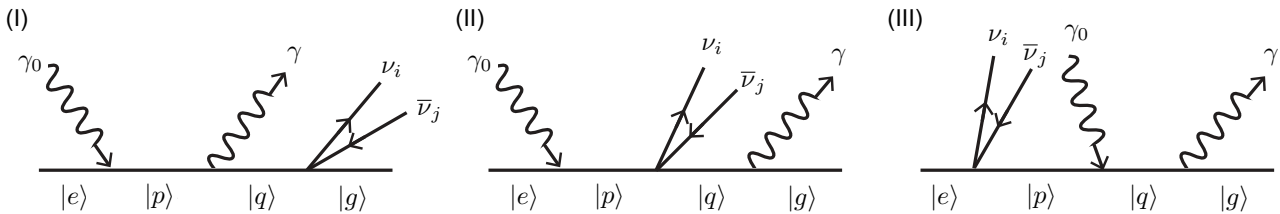}} \hspace*{\fill}
   \caption{
Three schemes of Raman-stimulated neutrino pair emission in de-excitation path,
$|e \rangle \rightarrow | p\rangle \rightarrow | q \rangle \rightarrow | g\rangle $.
}
   \label {three schemes}
 \end{center} 
\end{figure*}

Macro-coherent, double resonant Raman-stimulated neutrino pair emission (RANP) \cite{ranp} has
three possible schemes which differ in where neutrino pair is emitted, as depicted in Fig(\ref{three schemes}).
In all three schemes  the neutrino pair $\nu_i\bar{\nu}_j$ (or simply $(ij)$) emission has a vertex given by
\begin{eqnarray}
&&
\langle b | \vec{S} \cdot \vec{{\cal N}}_{ij} e^{ i ( \vec{p}_1 +\vec{p}_2 )\cdot \vec{x}} | a \rangle
\,,
\label {nu-pair emission amp}
\end{eqnarray}
where $\vec{S}$ is the total 4f$^n$ electron spin operator of lanthanoid ions, and
$\vec{p}_i\,, i = 1,2$'s are momenta of emitted neutrinos.
Neutrino pair current amplitude $ \vec{{\cal N}}_{ij} $ is readily calculable \cite{renp overview}
in the standard electroweak theory assuming finite neutrino masses of either Dirac or Majorana type.
Rate calculation proceeds, first taking square of the amplitude (\ref{nu-pair emission amp}) and summing over
non-detectable neutrino helicities and momenta.
This neutrino variable integration can be done 
independently of atomic part calculation. We assume the long wavelength
approximation in the plane wave factor, the pair momentum  being of order eV range or less, hence
 $ |\vec{p}_1 +\vec{p}_2  | \ll $ inverse of atomic size typically of order keV.
With $e^{ i ( \vec{p}_1 +\vec{p}_2 )\cdot \vec{x}} =1 $,
one is  led to the neutrino factor (later given by eq.(\ref{2nu integral})\,)
 times atomic matrix element  $| \langle b | \vec{S}  | a \rangle |^2$ for unpolarized ion
targets we assume hereafter.
Electromagnetic vertex responsible for photon emission and absorption is thus dominantly of magnetic
dipole (M1) type, and satisfies  the selection rule of atomic quantum number change, $ | \Delta J | \leq 1 $.

\begin{figure*}[htbp]
 \begin{center}
 \epsfxsize=0.6\textwidth
 \centerline{\includegraphics[width=7cm,keepaspectratio]{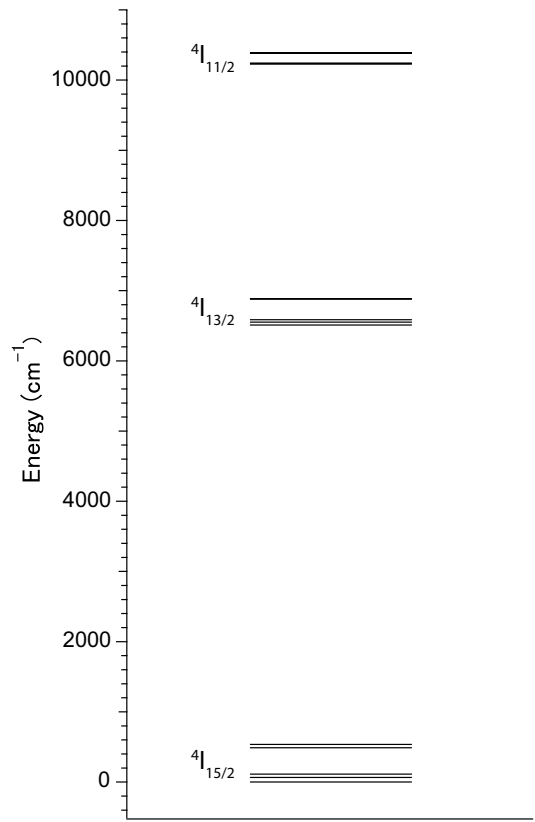}} \hspace*{\fill}
   \caption{ Three lower $J-$ manifold energy level diagram of Er$^{3+}:$Cs$_2$NaYF$_6$.
}
   \label {er3+ stark levels}
 \end{center} 
\end{figure*}

For atomic de-excitation path $|e \rangle \rightarrow | p\rangle \rightarrow | q \rangle \rightarrow | g\rangle $
of Fig(\ref{three schemes})
we assume  $ | \Delta J _{ab}| \leq 1$ for all atomic transitions, $|a \rangle \rightarrow | b\rangle $,
in order to maximize RANP rates.
In many lanthanoid ions energies of lower  $J$-manifolds  increase according to 
either decreasing or increasing $J$ values, as exemplified in Fig(\ref{er3+ stark levels}) for Er$^{3+}$.
We thus use three lowest $J$-manifolds to satisfy the selection rule $ | \Delta J _{ab}| \leq 1$
at each transition.
In \cite{ranp 2} we studied Sm$^{2+}$ ion at inversion center of host crystals, which requires
challenging infrared lasers.
In the present work we examine trivalent lanthanoid ion Er$^{3+}$ of different level spacing doped at inversion center of Y 
in Cs$_2$NaYF$_6$ and Y$_2$O$_3$, \cite{stark levels er3+}, \cite{er3+}, \cite{Er3+Y2O3},
which has O$_h$ crystal point group symmetry.
100\% doping (or replacement) gives Er$^{3+}$ number densities,
$5.40\times 10^{21}$cm$^{-3}$ for Cs$_2$NaYF$_6$ and $2.72 \times 10^{22}$cm$^{-3}$ for Y$_2$O$_3$.

As discussed in \cite{ranp} and \cite{ranp 2},
the squared mass function ${\cal M}^2$ plays crucial roles both
in the criterion of  dominant QED background rejection, namely McQ3 of \cite{mcqn},
and in RANP angular spectrum.
The function at resonances of Raman trigger $\gamma_0$ of energy $\omega_0$ and scattered $\gamma$
of energy $\omega$ is defined by ${\cal M}^2 = (p_{eg} + k_0 - k)^2\,, p_{eg} = (\epsilon_{eg}, r \epsilon_{eg},0,0) $
($\vec{p}_{eg} = r \epsilon_{eg} \vec{e}_x$  with $ \vec{e}_x$ the unit vector
along the excitation axis is the spatial phase vector imprinted at excitation).
The squared mass function is thus a function of two angles ${\cal M}^2(\theta_0, \theta)$, the Raman trigger direction $\theta_0$
and scattered direction $\theta$ both measured from the excitation direction.
The phase magnitude parameter  $r$ depends on excitation scheme one adopts,
and is unity, $r=1$, for a single  laser or two-laser excitation along the same direction.
From the reason stated below we fix the scattered direction along the excitation axis, hence $\theta = 0$ or $\pi$.
Under this circumstance,
\begin{eqnarray}
&&
{\cal M}^2(\theta_0, 0) = 4 \omega_0 (r\epsilon_{eg} - \omega ) \sin^2 \frac{\theta_0}{2}
+ 2 (1-r) \epsilon_{eg} \left( \omega_0 -  \omega + \frac{ 1+ r}{2} \epsilon_{eg}
\right) 
\,.
\label {msq f}
\end{eqnarray}
RANP experiments should be conducted at trigger directions when the quantity
${\cal M}^2(\theta_0, 0)$ is positive, with its values close to neutrino pair masses $(m_i + m_j)^2$.
As in \cite{ranp} and \cite{ranp 2}, we take the double resonance scheme to fix $\omega_0$ and $\omega$
for maximal RANP rates.
Scheme \lromn1 gives $\omega_0 = \epsilon_{pe} $ and  $\omega = \epsilon_{pq} $,
while Scheme \lromn2  $\omega_0 = \epsilon_{pe} $ and  $\omega = \epsilon_{qg} $.

It is difficult to satisfy the macro-coherence condition, namely the momentum and the energy conservation
given by 4-momentum notation, $p_{eg} + k_0 = k + p_1 + p_2 $ with $p_i$ two
neutrino energy-momentum, near the excitation axis for $r=1$.
As seen from eq.(\ref{msq f}), ${\cal M}^2(\theta_0, 0) $ tends to be negative  for $r > 1$,
hence we take $0< r < 1$ except in the case of $\theta=\pi$ later.
For counter-propagating two-laser irradiation for excitation two excitation photons
have energies of $(1 \pm r) \epsilon_{eg}/2 $.
In all paths discussed below we first verify the positivity of squared mass function ${\cal M}^2(\theta_0, 0) >0$
to ensure absence of McQ3 background.

\begin{figure*}[htbp]
 \begin{center}
 \epsfxsize=0.5\textwidth
 \centerline{\includegraphics[width=6cm,keepaspectratio]{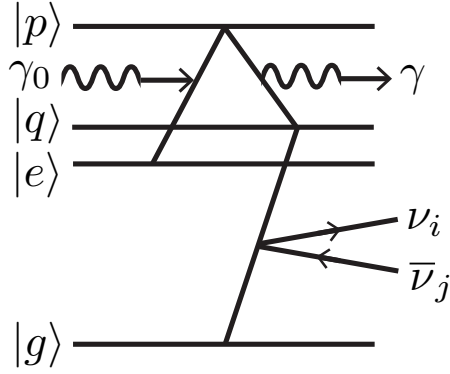}} \hspace*{\fill}
   \caption{ 
Energy level diagrams indicating  absorption and emission of photons and a neutrino-pair
corresponding to scheme \lromn1.
}
   \label {drenp levels}
 \end{center} 
\end{figure*}

\vspace{0.5cm} 
\section
{\bf Raman stimulated neutrino pair emission rate}

In the third order of perturbation theory which regards four-Fermi interaction
as a primary hamiltonian,
the squared amplitude in scheme \lromn1 is given by
\begin{eqnarray}
&&
| {\cal A}|^2 = \frac{ |H^{\gamma}_{pe }  H^{\gamma}_{pq} H^W_{qg }|^2 }
{\left( (\omega_0 - \epsilon_{pe} )^2 + \gamma_1^2/4 \right)  
\left( ( E_1 +E_2 - \epsilon_{qg} )^2 + \gamma_2^2/4 \right)}
2\pi \delta (\epsilon_{eg} + \omega_0 - E_1 -E_2 - \omega )
\,,
\end{eqnarray}
ignoring irrelevant terms.
A double resonance occurs at $\omega_0 = \epsilon_{pe}$ and $E_1 + E_2 =  \epsilon_{qg}$ (equivalent to
$\omega = \epsilon_{pq} $).
Dominant electromagnetic vertex $H^{\gamma}_{ab}$
among 4f$^{11}$ $J$-manifolds is of magnetic dipole type with a small admixture of electric dipole induced
by crystal field effects as first pointed out in \cite{van vleck}.

Differential rate for unpolarized targets is calculated, by using the formula of \cite{ranp 2}, as
\begin{eqnarray}
&&
\frac{d\Gamma_{ {\rm RANP} }}{d\Omega} = 
288 \pi
\frac{\gamma_{pe}  \gamma_{pq} G_F^2}{\epsilon_{pe}^2\epsilon_{pq}^2 } 
\frac{n^4 V }{\Delta \omega_0 (\gamma_e^2 + \gamma_q^2 )^{3/2} }  \,
\eta  \sum_{ij}  F_{ij} ( \theta_0)
\Theta \left( {\cal M}^2  (\theta_0,  0) - (m_i +m_j )^2\right) 
\,,  
\label {diff rate0}
\\ &&
\hspace*{-0.5cm}
 F_{ij} (\theta_0) =  
\frac{1}{8\pi} 
\left\{
\left(1 - \frac{ (m_i + m_j)^2}{  {\cal M}^2 ( \theta_0, 0)} \right)
\left(1 - \frac{ (m_i - m_j)^2}{  {\cal M}^2 ( \theta_0, 0)} \right)
\right\}^{1/2} 
\left[ \frac{1}{2} |b_{ij}|^2
\left({\cal M}^2 ( \theta_0, 0) - m_i^2 - m_j^2  \right)
- \delta_M \, \Re( b_{ij}^2 ) \,m_i m_j \right]
\,,
\nonumber \\ &&
\label {2nu integral}
\\ &&
b_{ij} = U_{ei}^*U_{ej} - \frac{1}{2} \delta_{ij}
\,,
\label {squared mass f}
\end{eqnarray}
where $n$ is the number density of excited target ions.
The angular acceptance factor $d\Omega$ shall be estimated later when we discuss 
an experimental layout.
From the macro-coherence condition of energy-momentum conservation the squared mass function
is equal to the invariant squared mass of the neutrino-pair system:
${\cal M}^2(\theta_0,  \theta) = (p_1+p_2)^2 $.
The $3 \times 3$ unitary matrix $(U_{ei} )\,, i = 1,2,3,$ refers 
to the neutrino mass mixing \cite{pdg}.

The parameter $\eta$ is defined by laser power divided by $|\rho_{eg}|^2 \epsilon_{pe} n/2 $
where $\rho_{eg} = |c_e^* c_g|^2$ with $c_a$ the probability amplitude of state $|a \rangle$
in a purely quantum mechanical system without dissipation.
In actual system $\eta$ is time varying $ \eta(t)  $, and may be calculated using 
the Maxwell-Bloch equation, a coupled system of non-linear and partial differential equations
for fields and density matrix elements.
Systematic calculation of  $ \eta(t)  $ is beyond the scope of the present work,
but its sample calculations are given in  \cite{renp overview}.
We introduce a time independent  $\eta$  in the present work.

Dependence on $\propto n^4 V $ of RANP rate (\ref{diff rate0}) is understood as follows.
Rate of Raman stimulated resonant weak process is given by a product of Raman rate times weak process rate multiplied by
the lasting lifetime of intermediates state,  
hence $d\Gamma_{RW} = 4 d\Gamma_R d\Gamma_W/\sqrt{\gamma_e^2 + \gamma_q^2}$ \cite{ranp 2}.
Macro-coherence gives $n^2 V(2\pi)^3  \delta (\vec{p}_{eg} + \vec{k}_0 - \vec{k} -\vec{p}_1 - \vec{p}_2 )$
with the momentum conservation for the entire process. 
With the resonance dictated by trigger laser at  frequency $\omega_0 = \epsilon_{pe}$,
the  momentum conservation is maintained in Raman and weak processes simultaneously, 
$\vec{p}_{eg} + \vec{k}_0 = \vec{k}  $ and $\vec{p}_1 + \vec{p}_2 =0$,
which gives an extra $n$ dependence as shown in \cite{ranp}, \cite{ranp 2}.
Another $n$ factor arises from incident trigger laser power $E_0^2$ related to
definition or the $\eta$ factor.

According to \cite{er3+}, calculated radiative decay rates of 10\% doped Er$^{3+}$ in host Cs$_2$NaYF$_6$ are
\begin{eqnarray}
&&
^4{\rm I}_{13/2} \rightarrow ^4{\rm I}_{15/2}: ( 24.82^{{\rm MD}} + 2.46^{{\rm ED}})\, {\rm sec}^{-1}
\,,
\\ &&
^4{\rm I}_{11/2} \rightarrow ^4{\rm I}_{13/2}: (4.26^{{\rm MD}} + 0.42^{{\rm ED}})\, {\rm sec}^{-1}
\,, \hspace{0.5cm}
^4{\rm I}_{11/2} \rightarrow ^4{\rm I}_{15/2}: 4.13^{{\rm ED}}\, {\rm sec}^{-1}
\,.
\end{eqnarray}
MD(M1) and ED(E1) refer to magnetic and electric dipole transition rates calculated according to modified
Judd-Ofelt theory \cite{judd-ofelt}.
Squared atomic vertex (magnetic or electric dipole) 
amplitudes are equal to $3\pi \gamma_{ab}/\epsilon_{ab}^3$, 
which was used in RANP rate formula, eq(\ref{diff rate0}).
We further assume that radiative transition rates between two  specific Stark states in two different $J$-manifolds
are not much different for different Stark states.
Stark level splitting caused by crystal field has been calculated and compared with experimental data
in \cite{stark levels er3+}.
In the adopted de-excitation path we used experimental results obtained in this work.
For state $| q\rangle $ of  $^4$I$_{13/2}'$ manifold there are other choices of Stark levels:
 6586\,, 6552\,,  6510\,cm$^{-1}$'s, the last one being the case of Raman elastic scattering.

Non-radiative decay rates are expected to be large for higher energy states.
According to \cite{er3+} decay rate of $^4$ I$_{13/2}$ of lower energy is mainly radiative, but 
the next lower $^4$ I$_{11/2}$ is dominated by non-radiative rates
(non-radiative is ten times larger than radiative) at room temperature.
We need to experimentally investigate non-radiative decay rates $^4$ I$_{11/2}$ at lower temperatures
to examine how much radiative rates increase compared with non-radiative rates,
but in the present work we shall assume dominance of radiative $^4$ I$_{11/2}$ decay for simplicity.

We first consider the following de-excitation path 
of scheme \lromn1 for Er$^{3+}:$Cs$_2$NaYF$_6$, 

Path A: Er$^{3+}$,  $^4$I$_{13/2} \rightarrow ^4$I$_{11/2} \rightarrow ^4$I$'_{13/2} \rightarrow ^4$I$_{15/2}$

 $^4$I$_{13/2}$,  6510 cm$^{-1}$ (805.8  meV)\,, \hspace{0.2cm} 
 $^4$I$_{11/2}$,  10227 cm$^{-1}$  (1266 meV) \,, \hspace{0.2cm} 
 $^4$I$'_{13/2}$,  6883cm$^{-1}$ (852 meV) \,.

\begin{figure*}[htbp]
 \begin{center}
 \epsfxsize=0.6\textwidth
 \centerline{\epsfbox{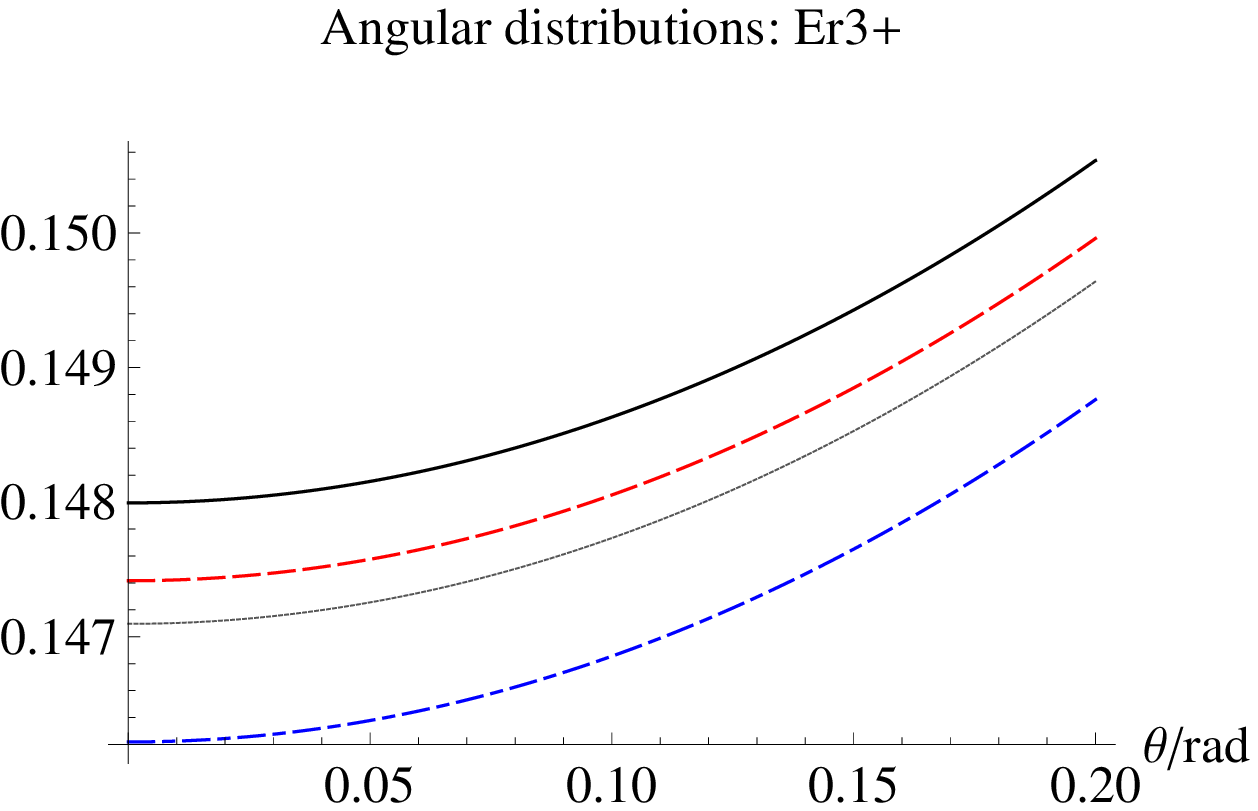}} \hspace*{\fill}
   \caption{ Raman trigger angular distribution given by $8 \pi  \sum_{ij} F_{ij} (\theta) $ in eV$^2$ unit,
assuming $r=0.8$ in scheme A:
Smallest neutrino mass of
$m_1 =5\,, 20 $meV's of Dirac NH in solid and dotted blacks,
and Majorana NH  in dashed red and dash-dotted blue. CP violating phases are assumed to vanish.
Absolute rate is derived by multiplying   
$R \Delta \Omega\,, R = 5.2 \times 10^{7} \,  {\rm sec}^{-1}\,
(\frac{n }{ 10^{15} {\rm cm}^{-3}})^4 \frac{V }{10^{-2} {\rm cm}^3 } \frac{100 {\rm MHz}}{\Delta \nu_0} \, 
\eta$, with $R$ defined in eq.(\ref{ranp rate numerical}) and $\Delta \Omega $ the angular acceptance factor.
}
   \label {md angular distribution}
 \end{center} 
\end{figure*}

Relevant dipole moment values  in the formula (\ref{diff rate0}) are 
\begin{eqnarray*}
&&
\frac{\gamma_{pe}}{\epsilon_{pe}^2 } = 
\frac{4.68 {\rm sec}^{-1}} {  0.460^2 {\rm eV}^2} \sim 5.74 \times 10^{-13} {\rm eV}^{-1}
\,, \hspace{0.5cm}
\frac{ \gamma_{pq}}{\epsilon_{pq}^2 } \sim 1.79 \times 10^{-14} {\rm eV}^{-1}
\,,
\\ &&
 (\gamma_e^2 + \gamma_q^2 )^{3/2} \sim 1.6 \times 10^{-41}\, {\rm eV}^3
\,,
\end{eqnarray*}
in eV unit, using data of \cite{stark levels er3+} and \cite{er3+}.
Taking $\Delta \omega_0 = 2\pi \Delta \nu_0 = 2\pi \, 100 {\rm MHz}$ for the trigger laser width gives
\begin{eqnarray}
&&
\frac{d\Gamma_{ {\rm RANP} }}{d\Omega} = R \frac{ 8 \pi  \sum_{ij} F_{ij} (\theta)}{ {\rm eV}^2} 
\,, \hspace{0.5cm}
R = 
5.2 \times 10^{7} \,  {\rm sec}^{-1}\,
(\frac{n }{ 10^{15} {\rm cm}^{-3}})^4 \frac{V }{10^{-2} {\rm cm}^3 } \frac{100 {\rm MHz}}{\Delta \nu_0} \, 
\eta 
\,.
\label {ranp rate numerical}
\end{eqnarray}
For convenience we changed  the notation $\theta_0$ to $\theta$.
In the figures of angular distributions this eV unit is used, giving 
$8 \pi  \sum_{ij} F_{ij} (\theta) $ of order $0.1 \sim 0.01 \,$eV$^2$.
Hence differential RANP rates of path A are of order $10^{6}$ sec$^{-1}$ assuming the same 
$n,V,\Delta \nu_0$ values.

In Fig(\ref{md angular distribution}) 
we show the sensitivity to the smallest neutrino mass and to Majorana/Dirac distinction.
In all calculations of rates we assume CP violating phases to vanish (case of CP conservation)
for simplicity.
The largest threshold rises are at the paired neutrino mass of (12) and (33) where
we order neutrino masses as $m_1 < m_2 < m_3$.
Input neutrino mass for numerical rate calculations is the smallest neutrino mass $m_1$ and we determine
other masses using data of \cite{pdg} along with mixing angles.

\vspace{0.5cm} 
\section
{\bf Bragg fiber to reject remaining QED backgrounds}

Macro-coherently amplified QED backgrounds have been classified in \cite{mcqn}
and they are called McQn,  when $n$ number of photons are involved.
Due to the host crystal symmetry of inversion center, McQ4 rejection containing E1 transitions
is nearly complete, but McQ4 events caused by four vertexes of M1 operators are still large.
We shall use  a Bragg fiber to suppress extra two-photon emission replacing neutrino-pair
\cite{renp in bragg fiber}, \cite{renp in bragg fiber 2}.
The idea is that due to an effective transverse photon mass in a kind of photonic crystal  or wave guide
some of two-photon emission may be prohibited in the Bragg fiber, but
neutrino pair emission is not.
To determine how much suppression we gain by this device it is necessary to estimate McQ4 event rate
of Er$^{3+}$ doped crystals.

\begin{figure*}[htbp]
 \begin{center}
 \epsfxsize=0.6\textwidth
 \centerline{\epsfbox{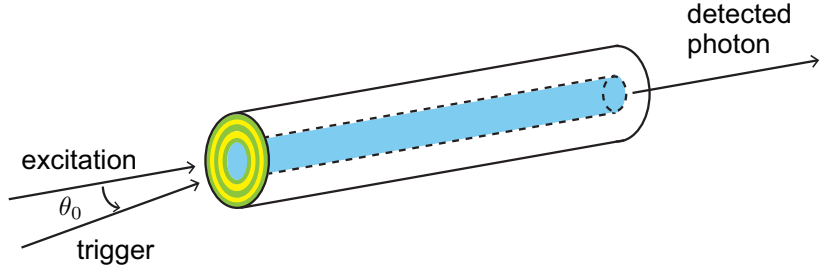}} \hspace*{\fill}
   \caption{ Illustration of an experimental layout.
Target crystal in light blue color is placed
 inside a hollow of Bragg fiber consisting of paired dielectric layers with a refractive index contrast.
Emitted and triggered photons can propagate only  along the Bragg fiber.
}
   \label {exp scheme}
 \end{center} 
\end{figure*}

\begin{figure*}[htbp]
 \begin{center}
 \epsfxsize=0.6\textwidth
 \centerline{\epsfbox{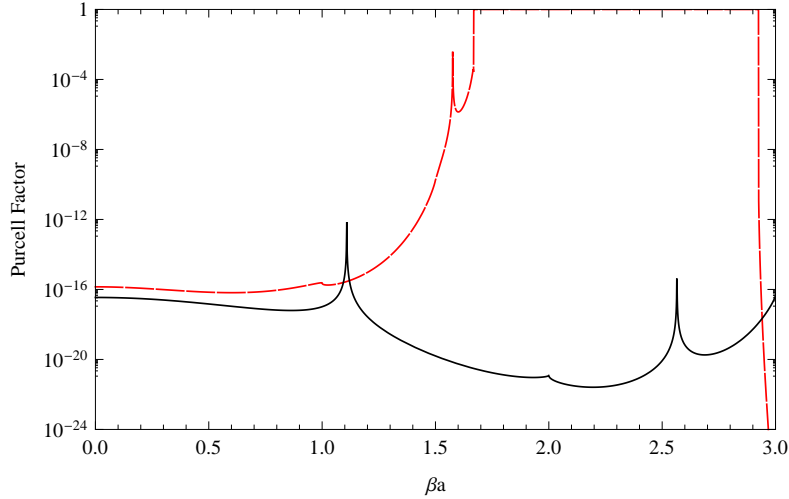}} \hspace*{\fill}
   \caption{ 
Purcell factor (ratio of intensities inside of  Bragg fiber to that in free space) vs $\beta$, 
the component of photon wavevectors along the fiber axis;
$\omega a=1$ in long-dashed red and $\omega a=2$ in solid black where 
$\omega$ and $a(=a_1+a_2)$ are, respectively, the frequency of the photon
and the thickness of one paired layer.
In the case of the two (single) photon emission between $|p\rangle$ and $|q\rangle$ states, 
each photon is  suppressed by a factor shown in  the red  $\omega a=1$ (the black $\omega a=2$) line.
Other parameters used in the calculation are: $a=2a_1=2a_2\simeq 11\;\mu$m,  $n_1=2$, $n_2=5$ and
the number of layers $N_p=35$.
Note that the maximum $\beta a$ in the case of the two (single) photon emission is $n_0= 1.5 \;(n_0\times 2=3)$ where $n_0$, the refractive index of target crystal, is assumed to be 1.5.
}
   \label {purcell factor}
 \end{center} 
\end{figure*}

In \cite{ranp 2} the ratio of McQ4 to RANP differential rates in scheme \lromn1 was calculated, to give
\begin{eqnarray}
&&
\frac{\Gamma_{{\rm McQ4} }(\theta)  }{ \Gamma_{{\rm RANP}}(\theta)} = 
\frac{ \Gamma_{2\gamma}(\theta) }{\Gamma_{2\nu}(\theta) }
\,,
\\ &&
\Gamma_{2\gamma} (\theta)= 
\frac{9\pi}{8} \frac{n}{\sqrt{\epsilon_{qg}^2 - {\cal M}^2(\theta)}}
\sum_x \,\frac{ \gamma_{xq}  \gamma_{xg} }{ \epsilon_{xq}^2 \epsilon_{xg}^2} 
\int_{\omega_-}^{\omega_+} d\omega \, \omega (\epsilon_{qg} - \omega )
\left( \frac{1}{\epsilon_{xq} + \omega} + \frac{1}{\epsilon_{xg} - \omega}
\right)^2
\,,
\label {2g rate}
\\ &&
\Gamma_{2\nu} ( \theta) = \frac{G_F^2 n}{2}  \sum_{ij}  F_{ij}  ( \theta) 
\Theta \left( {\cal M}^2 ( \theta) - (m_i +m_j )^2\right) 
\,,
\end{eqnarray}
with $\omega_{\pm} = \frac{1}{2} (\epsilon_{qg} \pm  \sqrt{\epsilon_{qg}^2 - {\cal M}^2(\theta)}) $.
We may approximate the summation over upper levels, $x$,  in eq.(\ref{2g rate})
by taking the lowest upper level, $x = p$.
The resulting single integral is calculated numerically, which turns out a slowly decreasing function of $\theta$, and
$\sim 1.6 $eV in the region of $\theta \leq 0.1$ in path A.
Using this value we obtain the ratio,
\begin{eqnarray}
&&
\frac{\Gamma_{{\rm McQ4} }(\theta)  }{ \Gamma_{{\rm RANP}}(\theta)} 
\sim 
\frac{18 \pi^2}{G_F^2 }
\frac{ \gamma_{pq}  \gamma_{pg} }{ \epsilon_{pq}^2 \epsilon_{pg}^2} \,\frac{ 1.6 {\rm eV}}{ \epsilon_{qg}}\,
\frac{{\rm eV}^2} { 8 \pi \sum_{ij}  F_{ij}  ( \theta) 
\Theta \left( {\cal M}^2 ( \theta) - (m_i +m_j )^2\right) }
\,,
\end{eqnarray}
with an approximation ${\cal M}^2(\theta) \ll \epsilon_{qg}^2$.
Numerically, the ratio is of order $5 \times 10^{21}$ times the inverse of  angular rate of order $10$,
giving of order $10^{22}$ ratio for Er$^{3+}$.

Since it is easier to forbid emission of lower energy photons  in Bragg fibers of realistic
layer thickness,
we calculate RANP angular distributions arising from de-excitation paths in scheme \lromn2
that use  smaller Stark level splitting within the same $J-$manifold.
As an example we use is

Path B: Er$^{3+}$:Cs$_2$NaYF$_6$ scheme \lromn2; \, $^4$I$_{15/2}$\, 537 cm$^{-1}  
\rightarrow ^4$I $_{13/2}$\, 6883 cm$^{-1} 
\rightarrow ^4$I$_{13/2}' $\, 6586\,, 6552\,, 6510 cm$^{-1}
\rightarrow ^4$I$_{15/2}$\, 0 cm$^{-1}$.

Explorable neutrino pair mass is limited by intra-$J$ manifold spacing $\epsilon_{pq}$.  
It is necessary to have trigger photons near the backward direction, $\theta_0 \approx \pi $ to the excitation axis.
With small $\epsilon_{eg} \sim 66.5$ meV,  Raman type of excitation with a large $r$ is required.
The value $r$ is chosen such that the center of mass system of the neutrino pair is at rest in the laboratory ($\vec{p}_1+\vec{p}_2=0$).
In this case, two photons by the McQ4 process have equal energy and thus background suppression by Bragg fiber becomes easier and efficient.
This way we derive $r=24.08$, and we calculate rates with this value.  
The angular acceptance factor $\Delta \Omega$ is estimated by placing a photon detector
at the end of cylindrical fiber, to give $\Delta \Omega \sim \pi d/l$ with $d$ the
cylinder radius and $l$ its length.

The necessary number of  paired dielectrics of refractive index contrast (2, 5)
is now worked out to search for the forbidden region of McQ4 events.
An example of suppression factors inside a Bragg fiber given by the Purcell factor \cite{photonic crystal}
is illustrated in Fig(\ref {purcell factor}).
This calculation uses an approximation that replaces the cylindrical layer configuration of Bragg fiber by the slab
configuration much easier for estimation.
This example may be used for two-photons  emitted nearly back to back in any direction with equal energy $\omega_*\,,
2\omega_* = \epsilon_{pq} $,
mimicking two-photon emission of total momentum zero.
The suppression factor of  two-photon emission with $\omega_* a $ =1 (red curve in Fig(\ref {purcell factor})\,)
is less than $10^{-24}$, sufficient to kill McQ4 events.

\begin{figure*}[htbp]
 \begin{center}
 \epsfxsize=0.6\textwidth
 \centerline{\epsfbox{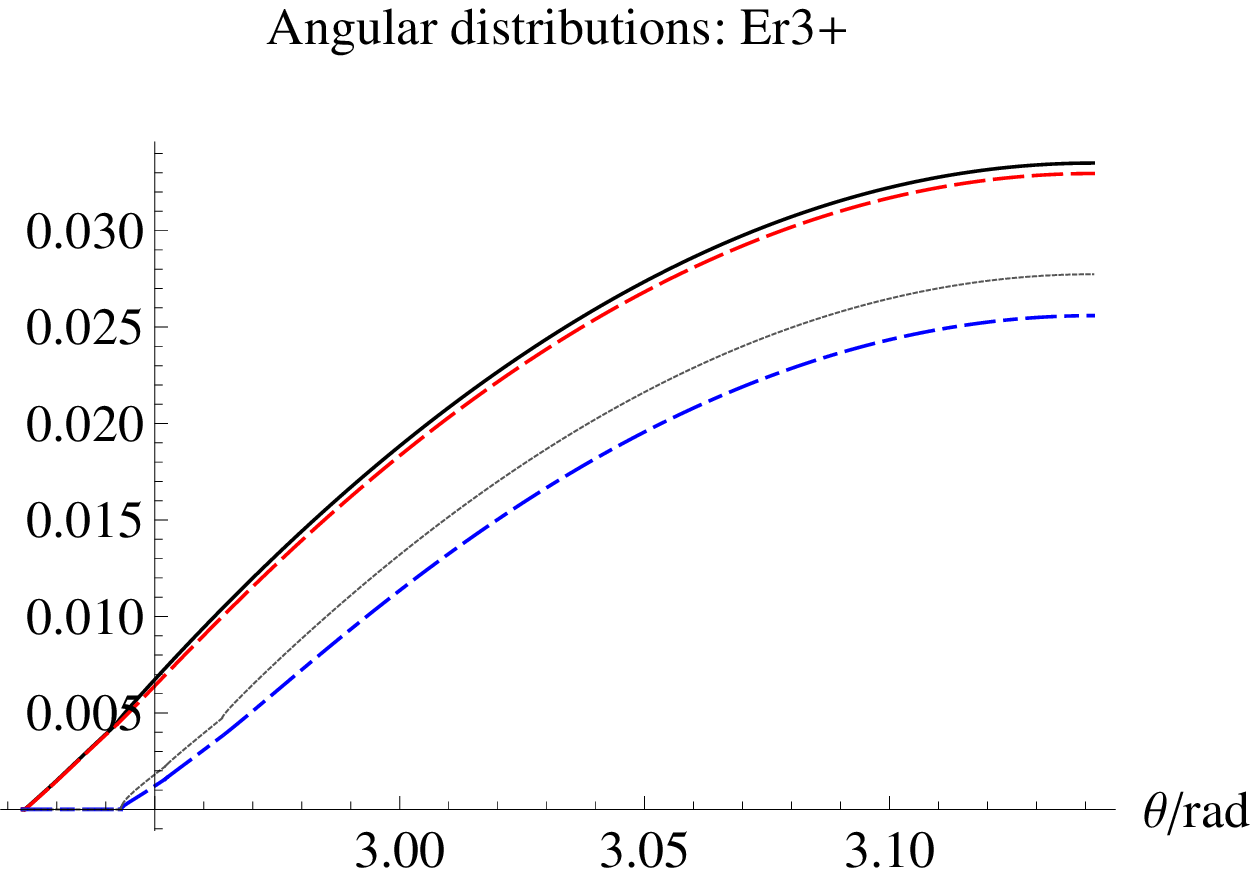}} \hspace*{\fill}
   \caption{ Raman trigger angular distribution  given by $8 \pi  \sum_{ij} F_{ij} (\theta) $ in eV$^2$ unit,
 assuming  different $r=24.08 $ in path B of Er$^{3+}$:Cs$_2$NaYF$_6$:
Dirac NH cases of smallest mass 5 meV in solid black, 50 meV in dotted black,
Majorana NH cases of smallest mass 5 meV in dashed red, and 50 meV in dash-dotted blue.
CP violating phases are assumed to vanish.
}
   \label {md angular distribution 2}
 \end{center} 
\end{figure*}

A different type of McQ4 event in which extra photons $\gamma_1 \gamma_2$ are emitted
between $|q \rangle $ and $|g \rangle $ of level spacing $\sim 800$ meV is inevitable in the kind of
Bragg fiber discussed here.
With a signal photon of energy $37 \sim 46$ meV suppressed by $< 10^{-16}$ 
 in the Bragg fiber, McQ4  rate is estimated $10^6$
relative to RANP rate.
Due to different event topology this McQ4 events are distinguishable from RANP signals, but
depletion in the state $|p\rangle$ gives effective reduction of RANP rate by $10^6$.
It is necessary to make simulations to compute depletion factor more precisely.
In Fig(\ref{md angular distribution 2}) we illustrate a result of angular distribution for this path.

\vspace{0.5cm} 
\section
{\bf Prospects}

An explorable range of neutrino mass is limited by
realistic fabrication of Bragg fiber  whose layer sizes are several tens of micron, inverse of 25 meV, at minimum.
Since one of the strongest thresholds is $(12)$ pair,
this limits the pair mass less than $\sqrt{m_1^2 + 10^2}+m_1 = 25$ meV, which corresponds to
the smallest neutrino mass $m_1 <$ 11 meV.
$( 3,3)$ threshold identification requires rejection of McQ4 extra photon of energy $\sim 60$ meV, which
is difficult to realize in the present scheme of Bragg fiber.

\begin{figure*}[htbp]
 \begin{center}
 \epsfxsize=0.6\textwidth
 \centerline{\epsfbox{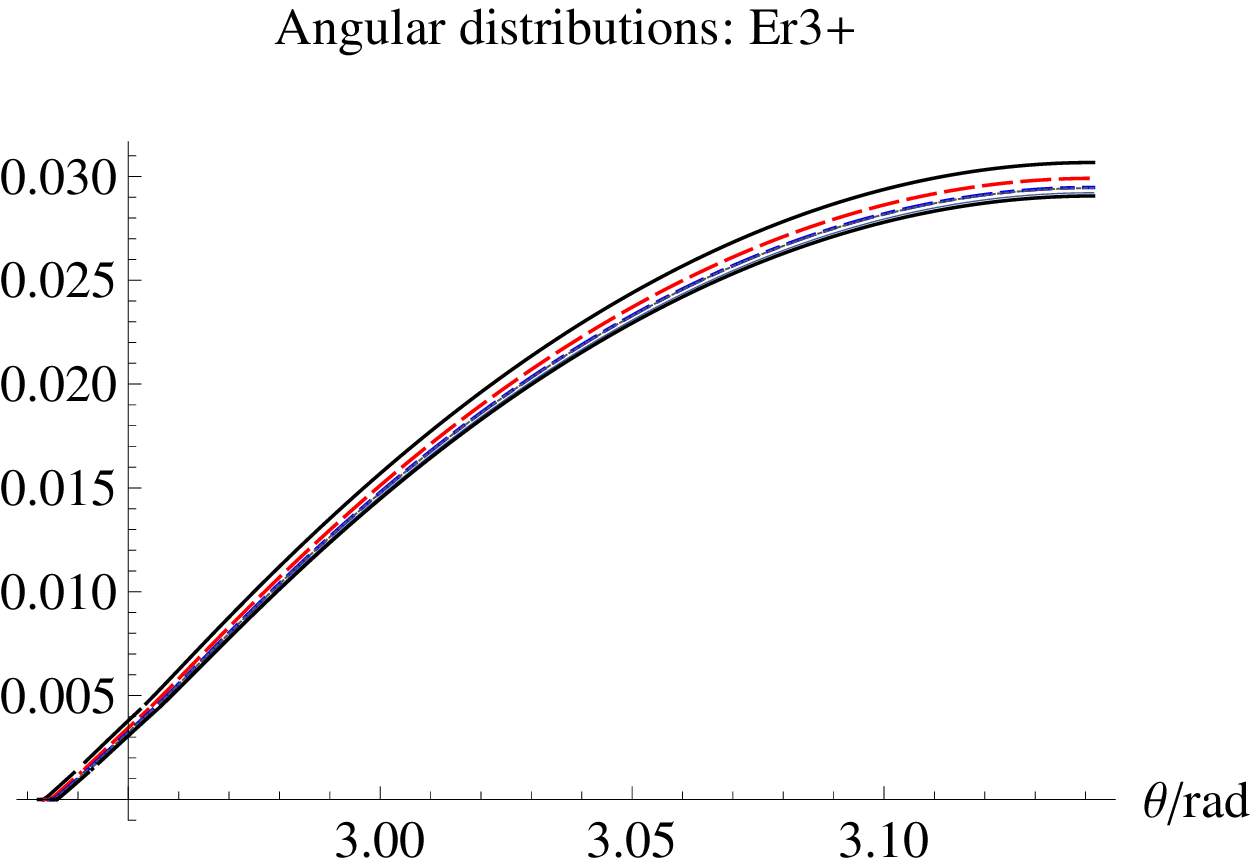}} \hspace*{\fill}
   \caption{ Angular distribution of RANP trigger photon given by $8 \pi  \sum_{ij} F_{ij} (\theta) $ in eV$^2$ unit
using Stark levels within $^4$I$_{13/2}$ manifold in scheme \lromn2 of Er$^{3+}$:Y$_2$O$_3$.
Smallest neutrino mass of
$m_1 =10\, $meV's of Majorana NH with $r= 24.08$. De-excitation path via
6510\,, 6542\,, 6588\,, 6594\,, 6684\,, 6840 \,, 6867 cm$^{-1}$ in increasing rate order.
}
   \label {yo distribution}
 \end{center} 
\end{figure*}

An equally interesting crystal is Y$_2$O$_3$ which can host Er$^{3+}$ at 100\%.
All seven Kramers doublets of Er$^{3+}\, ^4$I$_{13/2}$ in this crystal are identified with measured Stark level
energies: 6510\,, 6542\,, 6588\,, 6594\,, 6684\,, 6840 \,, 6867cm$^{-1}$  \cite{Er3+Y2O3}.
With a high resolution of detected photon energy it is in principle possible 
to measure six angular distribution curves via different paths as in Fig(\ref{yo distribution}).
Unfortunately, no radiative decay rates are not known for this crystal, and
one cannot predict RANP rates precisely.
Moreover, Er sites can be at either of two Y sites one of which has no inversion center,
and one needs to calculate decay rates in this situation.
It would however be interesting to pursue an experimental scheme using  ceramic or poly-crystal of this target
in a hollow of Bragg fiber.

\vspace{1cm}
 {\bf Acknowledgements}

We thank  F. Chiossi at Padova  for bringing  the paper \cite{er3+} to our attention
and M. Tanaka at Osaka for discussions on the Bragg fiber.
This research was partially
 supported by Grant-in-Aid 19K14741(HH), 19H00686(NS), 17H02896(AY) and 17H02895(MY)  from the
 Ministry of Education, Culture, Sports, Science, and Technology.

\end{document}